# Spinel nitride solid solutions: charting properties in the configurational space with explainable machine learning

Pablo Sánchez-Palencia,[1,2] Said Hamad,[3] Pablo Palacios,[1,4] Ricardo Grau-Crespo,[5] Keith T. Butler[6*]

[1] *Instituto de Energía Solar, ETSI Telecomunicación, Universidad Politécnica de Madrid, Ciudad Universitaria, s/n, 28040, Madrid, Spain.*

[2] *Departamento de Tecnología Fotónica y Bioingeniería, ETSI Telecomunicación, Universidad Politécnica de Madrid, Ciudad Universitaria, s/n, 28040 Madrid, Spain.*

[3] *Department of Physical, Chemical and Natural Systems, Universidad Pablo de Olavide, 41013 Sevilla, Spain.*

[4] *Departamento de Física aplicada a las Ingenierías Aeronáutica y Naval. ETSI Aeronáutica y del Espacio, Universidad Politécnica de Madrid, Pz. Cardenal Cisneros, 3, 28040 Madrid, Spain.*

[5] *Department of Chemistry, University of Reading, Reading RG6 6DX, United Kingdom.*

[6] *SciML, Scientific Computing Department, Rutherford Appleton Laboratory, Harwell OX11 0QX, United Kingdom.*

*E-mail: keith.butler@stfc.ac.uk

**ABSTRACT**

Ab initio prediction of the variation of properties in the configurational space of solid solutions is computationally very demanding. We present an approach to accelerate these predictions via a combination of density functional theory and machine learning, using the cubic spinel nitride $GeSn_2N_4$ as a case study, exploring how formation energy and electronic bandgap are affected by configurational variations. Furthermore, we demonstrate the utility of applying explainable machine learning to understand the crystal chemistry origins of the trends that we observe. Different configuration descriptors (Coulomb matrix eigenspectrum, many-body tensor representation, and cluster correlation function vectors) are combined with different models (linear regression, gradient-boosted decision tree, and multi-layer perceptron) to extrapolate the calculation of ab initio properties from a small set of configurations to the full space with thousands of configurations. We discuss the performance of different descriptors and models. SHAP (SHapley Additive exPlanations) analysis of the machine learning models highlights how values of formation energy are dominated by variations in local crystal structure (single polyhedral environments), while values of electronic bandgap are dominated by variations in more extended structural motifs. Finally, we demonstrate the usefulness of this approach by constructing structure-property maps, identifying important configurations of $GeSn_2N_4$ with extremal properties, as well as by calculating accurate equilibrium properties using configurational averaging.

# 1. INTRODUCTION

Computational methods play a major role in the exploration of the configurational space of solid solutions. Density functional theory (DFT)[1,2] is the most extended materials simulation technique at electronic-level, because of its accuracy and relatively reduced computational cost compared to other methodologies. However, the application of DFT simulations to evaluate the properties of the huge number of configurations of ion distributions in alloys is still a formidable task. In recent years, machine learning (ML) is attracting a lot of attention due to its capability to drastically accelerate materials simulations and reduce its cost by several orders of magnitude, acting as a support tool by using the results obtained from DFT programs.[3–9] A common criticism of ML approaches is that they represent black-box answers providing little physical insight, however recent work in explainable machine learning (commonly called XAI) shows the potential to extract meaningful explanations from complex ML models.[10,11] Herein we demonstrate the application of DFT combined with ML and XAI to exploring the configurational space of a prototypical mixed cation spinel material.

Silicon nitride, $Si_3N_4$, a wide bandgap ceramic with very high abrasion and oxidation resistance,[12,13] is one of the most commonly used 3:4 nitrides in industry. It can form different polymorphs, with the hexagonal $α$ and $β$ phases being the most energetically favorable ones. Advances in high-pressure and temperature synthesis methods have led to the discovery of the novel spinel (cubic) $γ$-phase of $Si_3N_4$ and analogous structures for a range of different cations, specially group 14.[14] This family of compounds possesses a promising combination of the properties of the more stable phases, plus outstanding mechanical features, with impressive hardness values,[15] and great tunability of their electronic properties.[16] Spinel phases of $Sn_3N_4$ and $Ge_3N_4$, with reduced bandgap values (1.6 eV and 3.5 eV, respectively),[17] in comparison with their respective $α$- and $β$-phases, are stable semiconductors with large exciton binding energies and electron mobilities.[18] This set of properties opens the door for new applications beyond coatings and mechanical applications, for example, in light emitting diodes, photocatalysis, sensors or electrodes for batteries.[17–20] Additionally, a recent study pointed to spinel compounds, with their uncommon structure with mixed tetrahedral and octahedral bonding, as potential candidates to substitute conventional absorber materials in solar cells, considering them a possible good compromise between typical high defect tolerance and long-term stability of tetrahedral and octahedral systems respectively.[21]

ML has made a considerable impact on the field of alloy and solid-solution materials design. Given the complex, high-dimensional search spaces, methods that can accelerate predictions and guide and inform choice of experiments are particularly promising. ML has been used to develop accurate, but computationally efficient interatomic potentials, that can be applied for calculating phase diagrams,[22] and for exploring high-entropy alloys.[23] ML has also been used, for example to explore the binding energies at the surfaces of alloy catalysts, a task that would be too computationally demanding using first principles approaches.[24] We previously demonstrated how ML could be used to predict band gaps and formation energies for arbitrary configurations of the ionic solid solution (Mg,Zn)O.[3] ML approaches are also extremely useful for guiding experimental studies and have been used for example to fuse experiment and theory for optimization of a halide perovskite solid-solution to optimize stability[25] and to discover optimal phase change memory materials in a complex composition/processing space.[26]

The present work explores Sn/Ge nitride solid solutions, which have been theoretically predicted to have favourable electronic properties for use in high-efficiency solar cells, such as tandem or intermediate-band solar cells.[27] Here, instead of attempting to engineer the properties of the solid solution via changes in composition, as done by Hart *et al.* for ternary Si/Ge nitrides,[28] we investigate how properties change as a function of the distribution of cations in a single composition, $GeSn_2N_4$. This is in spirit of growing theoretical and experimental work that has demonstrated that controlling the ion distribution is a viable route to tune the physical properties and functional behavior of materials.[29–33] In this work, we use a combination of DFT and ML techniques to investigate the properties of the $GeSn_2N_4$ solid solution as a function of cation distribution configuration. We focus on the mixing energies $E_{mix}$, and the electronic bandgaps $E_g$. Supervised learning of these properties, from DFT results obtained for a small fraction of the total number of configurations, allows their accelerated prediction for the full configurational space. We test different combinations of descriptors and models, with the goal of developing a methodology that allows us to perform a computationally efficient investigation of the properties of these spinel nitrides, and progress towards their optimization for photovoltaic applications. We then use the SHAP[34] (SHapley Additive exPlanations) approach to extract physical interpretations from our model predictions, linking observed changes in formation energy and bandgap to specific crystal chemistry motifs. We use our approach to construct structure-property maps that can be used to efficiently identify important and useful relationships and pick out extreme examples and can be combined to obtain ensemble averages, which highlights the potential application of processing conditions for property tuning. The methods that we present are applicable to a wide range of important problems where rational design of solid solutions is a key technological enabler.

## 2. METHODOLOGY

### 2.1 Density functional theory simulations

All the DFT calculations carried out in this work were done with the Vienna *Ab initio* Simulation Package (VASP),[35,36] following the projector augmented wave (PAW) formalism.[37] The results presented come from full structural relaxations of the different configurations, performed at a generalized-gradient approximation (GGA) level with the Perdew-Burke-Ernzerhof (PBE) functional.[38] We used Hubbard corrections for the *d*-orbitals of Sn and Ge, following Dudarev's approach (GGA+U)[39,40] with a $U_{eff}$ value of 2 eV, as this correction was found to improve the agreement with experiment for the binary nitrides, in terms of both geometric and electronic structure. For all the calculations, the plane wave kinetic energy cutoff was set to 520 eV, which is 30% higher than the suggested value for standard accuracy calculations with the given set of PAW potentials. For integration in the reciprocal space, the Brillouin zone has been sampled with a 2x2x2 k-points grid.

It is well known that the GGA generally produces underestimated bandgaps. In this work we found that the underestimation was ~1.5 eV against experimental results for the pure binary reference compounds. The GGA+U correction, although improving over GGA, was not enough to overcome this deficit. For that reason, and trying to produce physically and experimentally meaningful results for that central property, the screened hybrid functional by Heyd-Scuseria-Ernzerhof (HSE)[41] was also used for the reference binary compounds and subsequently for a subset of configurations within the calculated space. We will discuss below that a strong correlation exists between GGA+U and HSE bandgaps, which allowed us to reduce the computational cost of accurate bandgap predictions across the configurational space. Our HSE calculations, with 25% of exact exchange and a range-separation parameter of 0.2 (corresponding to the HSE06 parameterization), were single-point calculations on top of the GGA+U relaxed structures. **Table 1** presents the results of lattice parameter and bandgap for the reference binary compounds and the different calculation levels.

**Table 1** Calculated and experimental values of lattice constant and bandgap for the reference compounds γ-$Sn_3N_4$ and γ-$Ge_3N_4$.

| | Lattice constant (Å) | PBE+U bandgap (eV) | HSE bandgap (eV) (at PBE+U geometry) |
|---|---|---|---|
| γ-Sn$_3$N$_4$: DFT | 9.05 | 0.62 | 1.66 |
| Exp. | 9.03 | - | 1.6±0.2,[17] 1.6[18] |
| γ-Ge$_3$N$_4$: DFT | 8.24 | 2.30 | 3.50 |
| Exp. | 8.21 | - | 3.5±0.2,[17] 3.65±0.05[42] |

**2.2 Cation configurations**

We consider the distribution of Ge and Sn atoms in the 56-atom spinel cubic unit cell of γ-GeSn$_2$N$_4$. The full configurational space in that cell has a total of 4222 symmetrically inequivalent configurations, which were generated using the SOD code, together with their degeneracies in the full configurational space.[43] Out of those configurations, only 1013 (24%) were calculated at DFT level, to train and test the ML models. For the HSE calculations, a subset of 59 configurations was selected.

The configurations for DFT calculations were chosen randomly but we ensured that the selection was homogeneously distributed across all the different "inversion degrees" within the space. The inversion degree of an AB$_2$X$_4$ spinel is the fraction of tetrahedral sites that is occupied by B cations. Since the number of octahedral sites is double the number of tetrahedral sites in a spinel structure, a "normal" distribution is defined as one where the A cations occupy the tetrahedral sites and the B cations occupy the octahedral sites. The concept of inversion degree has been widely used for II-III$_2$-VI$_4$ spinels, where it represents the fraction of tri-valent cations in tetrahedral positions, and for II$_2$-IV-VI$_4$ spinels, where it represents the fraction of di-valent cations in tetrahedral positions. In our case, both cations are formally tetravalent, but the concept of inversion degree remains useful as a single, scalar descriptor of the cation distribution. The inversion degree, $y$, of GeSn$_2$N$_4$ is defined here as the fraction of tetrahedral sites occupied by Sn cations. Thus, $y=0$ refers to a normal or direct spinel, where all the Ge atoms are in tetrahedral positions, and all Sn are in octahedral positions, avoiding partial occupancy of sites; whereas $y=1$ refers to a fully inverse distribution where all the tetrahedral sites are fully occupied by Sn. The formula unit can then be written as (Ge$_{1-y}$Sn$_y$)[Sn$_{2-y}$Ge$_y$]N$_4$, where the round brackets "()" represent the tetrahedral sites and the square brackets "[]" represent the octahedral sites. Disaggregated numbers of total and calculated configurations for each of the inversion degrees are presented in **Table S1** of the supplementary material.

**2.3 Descriptors and models for machine learning**

A numerical way of fully describing structures is needed to use ML or other statistical models; these numerical representations are the descriptors. These descriptors must be invariant to symmetry operations, complete and unique, meaning being able to distinguish between any two different structures.[44]

Although some other descriptors like Ewald and Sine matrices[45] have also been tested in this work, only the main three descriptors for which we obtained best results are explained in detail here. The first one is the Coulomb matrix,[46] whose elements ($C_{ij}$) represent the electrostatic interactions between the nuclei in the compound:

$$C_{ij} = \begin{cases} 0.5 Z_i^{2.4} & \text{for } i = j \\ \dfrac{Z_i Z_j}{R_{ij}} & \text{for } i \neq j \end{cases}$$

where $Z_i$ is the atomic number of atom $i$, and $R_{ij}$ is the distance between atoms $i$ and $j$. For the sake of simplicity, the input for the machine learning model is not the Coulomb matrix directly, but the Coulomb matrix eigenspectra (CME), which is the vector formed by its eigenvalues sorted by size.

The second descriptor used is the many-body tensor representation (MBTR),[47] a structure descriptor that is easily interpretable and visualised. MBTR consists of a set of values of interatomic distances and angles around which weights and broadening are added, obtaining a set of spectra:

$$f_k = \sum_{i=1}^{N_a} \omega_k(i) \mathcal{D}(x, g_k(i)) \prod_{j=1}^{k} C_{Z_j, Z_{i_j}}$$

where $\omega_k$ and $\mathcal{D}$ represent the weighting and the broadening applied to the element matrix $C$, for every function $g_k$ considered which in our case define atom counts and inverse distances. The information represented is adjustable. For instance, dihedral angles could be included, but we have not considered them in our study, because of the extra complexity added to the descriptor, without a significant improvement in the results.

Finally, another descriptor with a long record in the prediction of solid solutions thermodynamic properties, are the cluster correlation functions (CCFs).[48] These are the basis of the cluster expansion method, although in that methodology a linear approach is used, whereas in our study we will also consider non-linear models based on the CCFs. The alloy configuration is described through the occupation of the different positions in the crystal lattice, selected in specific clusters or arrangements up to a desired order (individual atoms, pairs, trios, etc.). The CCF $X_{m\alpha}$ for a configuration *m* and cluster $\alpha$ can be defined as the average of the product of the functions $\phi_{ms}$ (which here take values of 0 and 1 depending on the atom occupying the site *s* in configuration *m*) over all the clusters of type α:

$$X_{m\alpha} = \frac{1}{\Omega_\alpha} \sum_{\beta \equiv \alpha} \prod_{s \in \beta} \phi_{ms},$$

where $\Omega_\alpha$ are the multiplicities (number of symmetrically equivalent clusters of type $\alpha$ in the cell). In this work, we employed the Python package CELL[49] to obtain the CCF vectors $\mathbf{X}_m = \{X_{m\alpha}\}$ corresponding to clusters of up to third order for each of the 4222 symmetrically different configurations of $GeSn_2N_4$.

All these descriptors were used to feed different ML models, to test the performance of each descriptor-model pair, with the aim of finding the best fit for the configurational space we are investigating. Among those models we used a simple multilinear regressor (LR), including LASSO regularization[50] to constrain coefficients to physically reasonable values. Two more complex models were also trained: a gradient boosted decision tree regressor (GBDT),[51] and a multilayer perceptron (MLP) neural network.[52] The MLP employed is a class of feedforward neural network, in this case with a 5-layer architecture, with 256-128-64-32-1 nodes respectively, which was chosen based on a previous study of some of the authors of this work.[3] The main metric that has been used to assess the performance of the different models and descriptors is the mean absolute error (MAE), although other additional metrics have been used for evaluation and to explain certain details, like the coefficient of determination ($R^2$) and the maximum error ($\varepsilon_{max}$). A standard procedure of set splitting in train-validation-test subsets was followed, with 80-10-10 percentages respectively, unless otherwise stated. To reduce dependency on random initial weights of the models (for the case of non-linear models) and the specific subset used to train the model, an ensemble configuration with model averaging among 10 runs, with reshuffling of the trainset between consecutive runs, was used. That way, scoring of the performance has been evaluated with predictions averaged over the 10 runs. All models, data and the code needed to reproduce our results are available in an open code repository; see data availability statement for details.

## 3. RESULTS

### 3.1 Prediction of HSE bandgaps from GGA+U values

We first examine whether we can use the GGA+U bandgaps to predict the HSE values, which are expected to be in much closer agreement with experiment, as seen in **Table 1** for the binary compounds. We first perform a simple linear regression between the two sets of bandgaps for the 59 selected structures, which produces the model:

$$E_g^{HSE} = 1.050 \cdot E_g^{GGA} + 1.041 \text{ eV}.$$

Although this is a good model (**Fig. 1a**), with a mean absolute error (MAE) of 7.5 meV calculated using 10-fold cross validation, the model can be improved substantially by including the inversion degree (*y*) as an additional parameter (**Fig. 1b**), which leads to the regression line:

$$E_g^{HSE} = 1.082 E_g^{GGA} - 0.047 \text{ eV} \times y + 1.045 \text{ eV},$$

with a 5-times smaller MAE (1.5 meV) from 10-fold cross validation, that is below the typical precision of DFT calculations. Given the high accuracy of the GGA+U → HSE bandgap transformation, in what follows we will use the transformed bandgaps for training of the ML models. In this way, all the bandgaps reported here can be considered at HSE level.

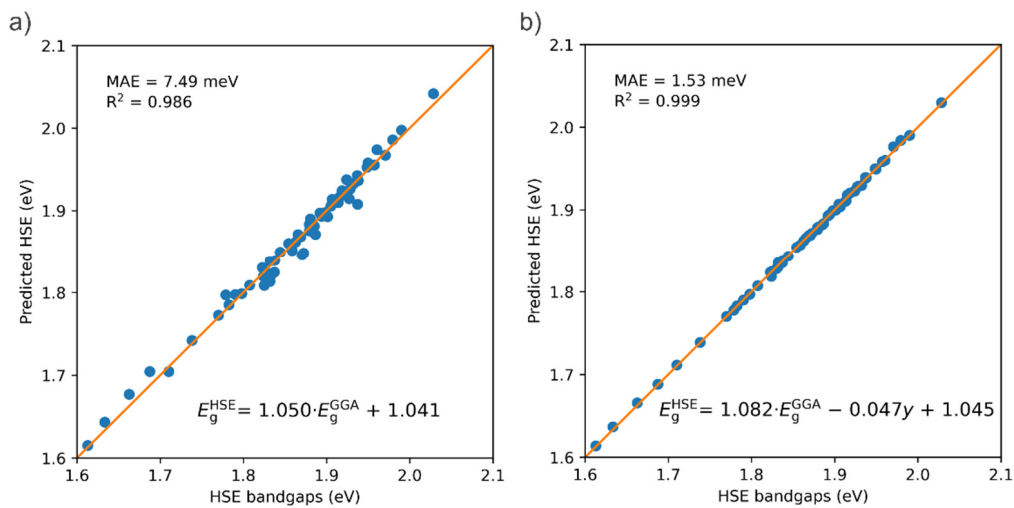

**Fig. 1** Performance of two models to predict the HSE bandgaps from the GGA+U bandgaps: a) simple linear model not using the inversion degree; and b) bi-linear model using the inversion degree *y* as additional parameter.

### 3.2 Machine learning models

The LR, GBDT and MLP models were first trained using 810 configurations (80% of the DFT-calculated dataset), and the three types of descriptors (CME, MBTR and CCFs), for predicting both mixing energy $E_{mix}$ (defined as the total energy of a particular configuration minus the total energy of the corresponding binary compounds weighted by the relative concentration of their respective cations in the configuration) and bandgap $E_g$. The MAEs obtained for the previously unseen test set are presented in **Table 2.**

**Table 2.** Mean absolute errors (MAE) of mixing energy ($E_{mix}$) and bandgap ($E_g$) predictions for the test set with different descriptors and models. Values highlighted in bold correspond to the best performing model-descriptor pair for each property.

| $E_{mix}$ MAE (meV) | | Descriptors | | |
|---|---|---|---|---|
| | | CME | MBTR | CCF |
| Models | MLP | 30 | 29 | 6 |
| | GBDT | 25 | 34 | 13 |
| | LR | 30 | 16 | **3** |

| $E_g$ MAE (meV) | | Descriptors | | |
|---|---|---|---|---|
| | | CME | MBTR | CCF |
| Models | MLP | 15 | 16 | **6** |
| | GBDT | 11 | 17 | 9 |
| | LR | 13 | 15 | 8 |

The CCFs descriptor outperforms the CME and MBTR, not only for mixing energies, which could be expected due to the additive character and linearity of this property, but also for bandgaps, where historically CCF-based methods have found more problems to predict the correct behavior.[53] Interestingly, this is the opposite of what was found in the previous study by Midgley *et al.*,[3] where CME showed much better results than CCFs for the mixing energy and bandgap prediction in the configurational space of an (Mg,Zn)O solid solution. Further research including different alloys is needed to investigate how and why the best descriptor depends on the nature of the alloy system. For the moment, we suggest that the nature of the chemical bonding in the different systems might play an important role. The more local character of CCFs, which describes mainly short distance arrangements of atoms through clusters, might be more suitable for covalent systems like this spinel, whereas the CME better captures the long-range interactions in a more ionic system like the (Mg,Zn)O solid solution.

The LR model is the best for describing mixing energies, at least with the local descriptors, MBTR and CCFs, reflecting the fact that energies are additive with respect to local contributions.

However, when predicting bandgaps with the CCFs descriptor, the non-linearity of the MLP neural network permits it to outperform the LR model. The small difference between models might justify the use of simpler linear models in some cases, when training time and resources could be a limitation or a concern. Beyond the difference in the descriptor, the results of the best performing methods are very similar to those obtained for (Mg,Zn)O in Ref. 3, considering the spread of the values for each system, and are accurate enough to be used confidently, being the best test MAEs equal to 3 and 6 meV for mixing energies and bandgap values respectively.

Different models can show large variations in performance depending on training data size; we now look at this effect. For this analysis, we focus on CCFs as the best descriptor and reduce the size of the training set to 505 or 202 configurations (50% and 20% of the total number of DFT-calculated structures, instead of the 80% used previously). The results of these tests are displayed in **Fig. 2**. The percentages presented there are referenced to the total number of configurations in the configurational space, corresponding to 5%, 12% and 19%, approximately. From there it can be clearly seen that, even with those reduced percentages, MLP and LR perform almost with the same accuracy, although GBDT drastically drops in its predictions, especially for the smallest set. Trends in here suggest that using bigger training sets could improve MLP performance, even with MAE values below those showed by LR for mixing energies, but at an increased cost that might not be worth it considering the already good results obtained with LR.

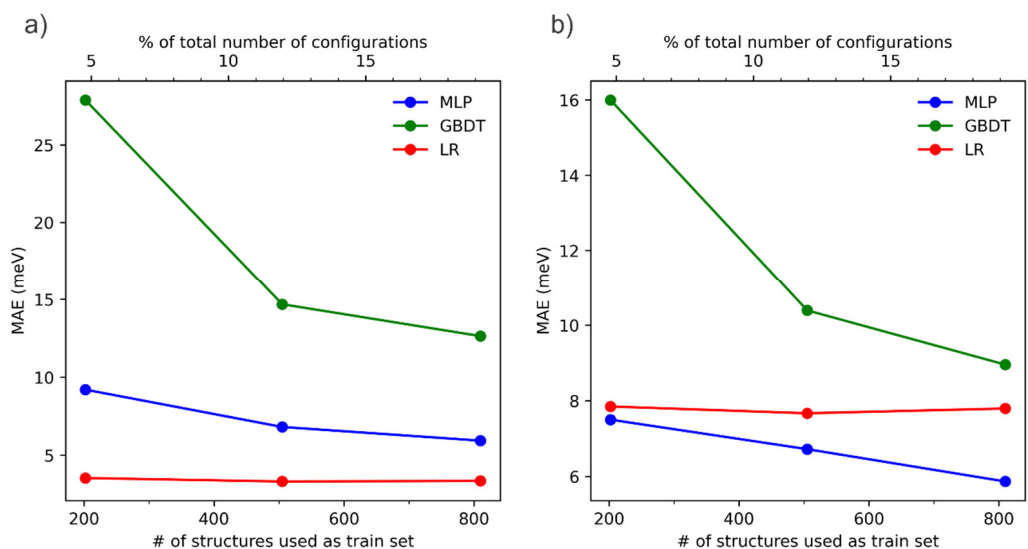

**Fig. 2** Mean absolute errors evolution upon train set size in the predictions of a) mixing energies and b) bandgaps with CCFs as descriptor and for the different models tested.

**Figs. 3a and 3b** present a direct comparison between calculated and predicted values of the test set, according to the best performing model-descriptor pair for each property. The high accuracy for mixing energies is clearly noticeable with a $R^2$ value close to unity. For bandgaps the metrics are also very good, with maximum errors below 30 meV.

Despite these positive results, it can be seen that errors in the prediction of bandgaps are mainly concentrated in extremal configurations, out of the energy range where most configurations are located and for that reason where the model is not as accurate in its predictions. That issue can be studied in more detail by plotting both calculated and predicted values against the inversion degree of the configurations, as shown in **Figs. 3c and 3d**. While for mixing energies there are no noticeable differences between calculations and predictions, for the case of bandgaps there are some poorer predictions. For example, the only configuration with inversion degree $y=0$ is one of the worst predictions within the test set, with the bandgap being significantly overestimated. Also, the configuration with the lowest calculated bandgap, with inversion degree $y=0.5$ is notably underestimated. In any case, the errors are still small and at worst of a few tens of meV. Clearly, the CCF-based models are excellent to make predictions in the configurational space of this solid solution.

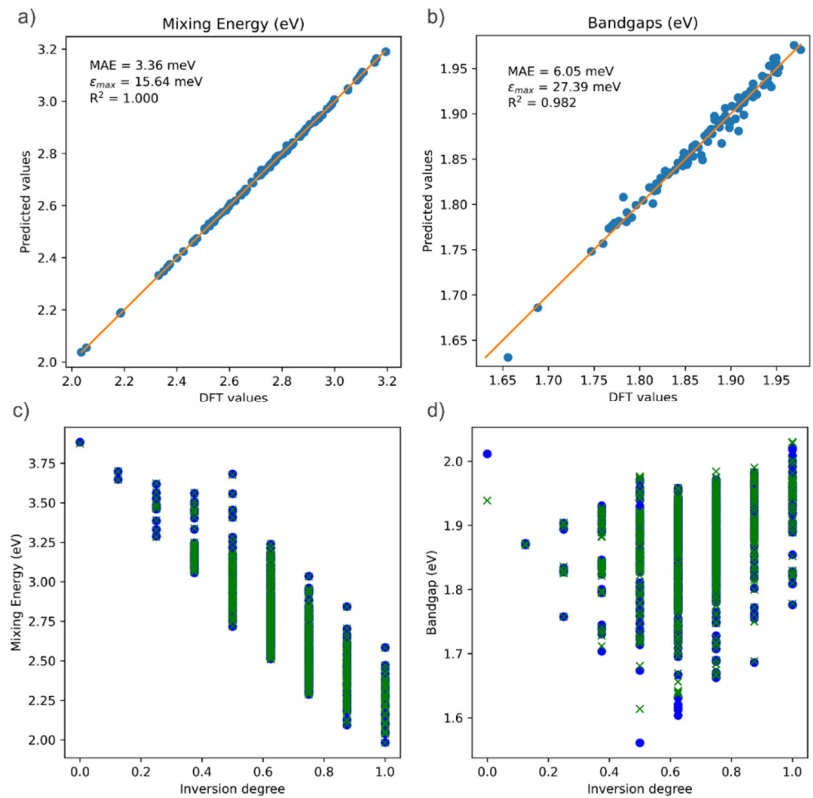

**Fig. 3** Correlation between ML-predicted and DFT-calculated values in the test set for a) mixing energies and b) bandgaps, including detailed metrics of performance. Also, ML-predicted (blue dots) and DFT-calculated (green crosses) values of c) mixing energies and d) bandgaps plotted against the inversion degree of the configuration.

### 3.3 Feature importance analysis

The absolute values of the coefficients of the linear models provide a measure of the importance of every descriptor (cluster) for predictions. For non-linear models, more advanced tools based on game theory, implemented on the SHAP Python library, are used with the same purpose. Applied to CCFs, we can see which kinds of clusters are those with more influence on the different results of both properties. **Fig. 4a** presents the linear coefficients for the different clusters within the LR model for predicting mixing energies. The results of the SHAP analysis for both properties are also presented in **Figs. 4b and c**. Within our CCFs definition, clusters 1 and 2 correspond to the single-site clusters for the different coordination sites (octahedral or tetrahedral), clusters from 3 to 9 correspond to site-pair clusters, while the rest represent clusters of three sites. It is also significant that, within groups of the same order, the clusters with the lower index represent groups of atoms of greater proximity than those with a higher index in the same group. In **Figs. 4b and 4c**, the $k$ parameter stands for the order ($k$ = 1, 2 or 3) and the $l$ parameter (in Å) stands for the maximum distance between atoms of the cluster. Details for all clusters in the expansion are given in **Table S2** of the supplementary material.

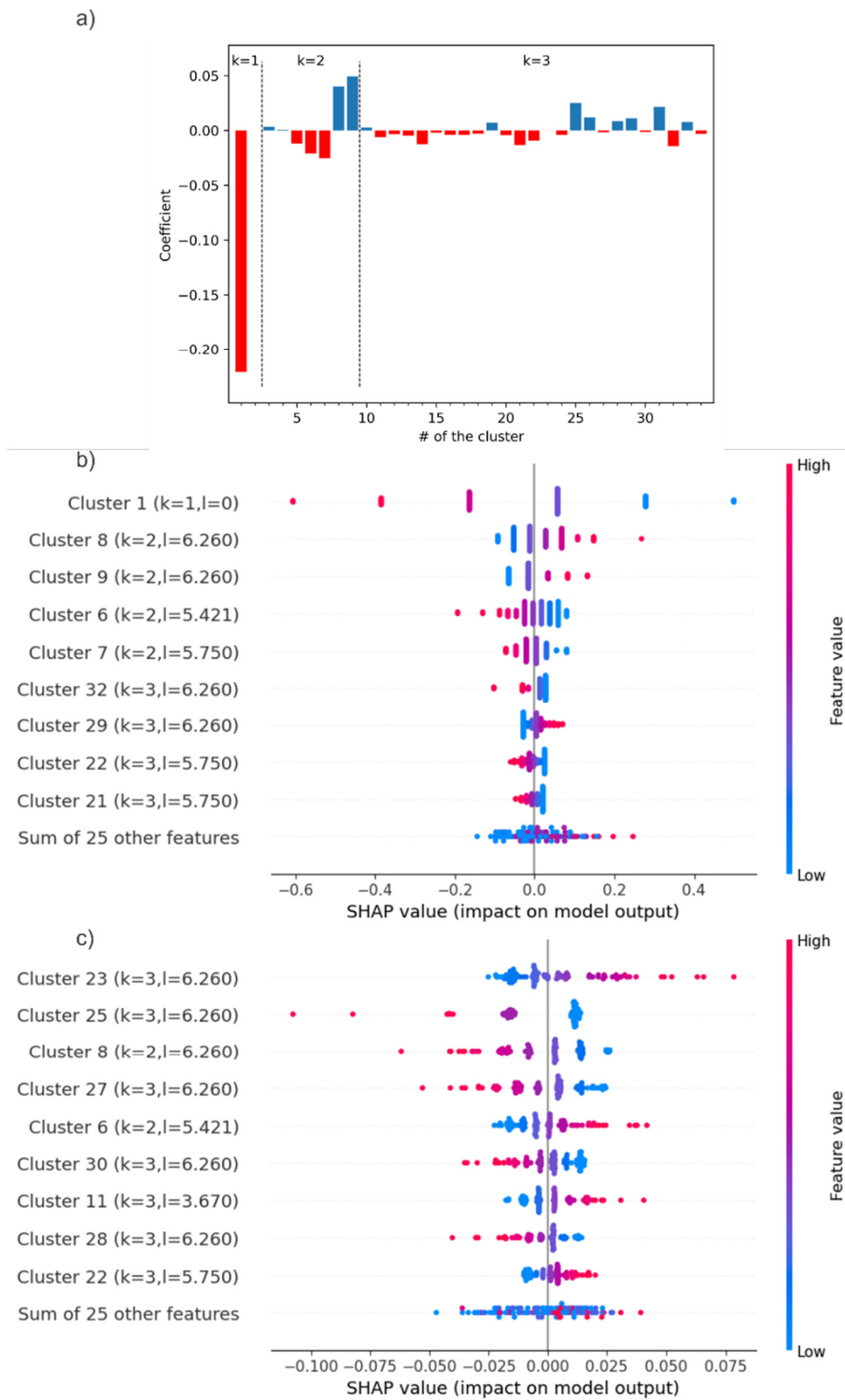

**Fig. 4** a) Linear coefficients for every cluster (for orders k=1, 2, and 3) in the model used to predict mixing energies. SHAP analysis of the impact of the different clusters within the CCFs on the predicted values of b) mixing energy and c) bandgap, where l (in Å) is the maximum distance between atoms of the cluster.

From the figures of the SHAP analysis and the linear coefficients presented in **Figs. 4a and 4b**, it can be seen that, when predicting mixing energies, the most important cluster is the single-site cluster for octahedral positions (cluster #1). Cluster #2 is the single-site cluster for tetrahedral positions, but because the total composition over the two sites is fixed, this cluster does not offer any additional degree of freedom and has formally zero importance. Subsequently, the next clusters in terms of relevance are clusters from 5 to 9, all clusters of order two. Thus, mixing energies are mainly defined by the inversion degree and by pair clusters to a lesser extent. This conclusion is consistent with the local character of the configuration energies, which, as previously said, have been traditionally described using CCFs within the cluster expansion method.

On the other hand, **Fig. 4c** shows that the most important clusters when predicting bandgaps are mainly clusters of order three. For predicting bandgaps, the specific arrangement of Sn and Ge atoms within the octahedral and tetrahedral lattice is thus much more important than for predicting mixing energies. It is interesting to see too that the difference between the most important clusters and the following ones is not as pronounced in bandgaps as in mixing energies predictions. The most important clusters involve mainly octahedral positions, although the effect of clusters involving tetrahedral positions is not negligible. This analysis is consistent with the bandgap being a global property of the ion distribution, rather than arising from the addition of local contributions (as in the case of the energy). Long-range interactions and atom arrangement patterns play a key role in determining the bandgap.

We can also gain additional physical insight to the performance of CCFs by investigating a covariance matrix of the different clusters (**Fig. 5**). Apart from clusters 1 and 2, there is no strictly redundant information because of the existence of direct or inverse correlation between cluster values. The correlation values between clusters are in general higher (in absolute value, meaning strong direct or inverse correlations) for pair clusters, where strong correlation between clusters involving tetrahedral and octahedral sites is noticed. However, that effect is much weaker in trios, which could be the reason for the higher complexity of bandgap predictions.

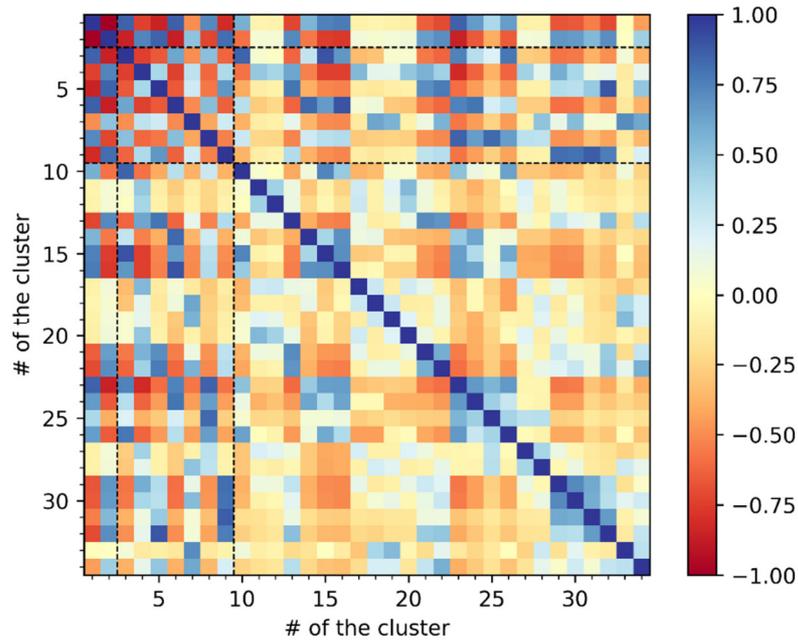

**Fig. 5.** Covariance matrix of the different values within the CCFs.

## 3.6 Applications of the predicted results for the full configuration space

We now illustrate the usefulness of being able to evaluate properties in the full configurational space, thanks to the ML model, which was trained with DFT data from just a small subset of that space. We will first discuss the identification of optimal configurations (e.g., the lowest-energy one, or the configuration with minimum or maximum bandgap), and then the calculation of equilibrium properties.

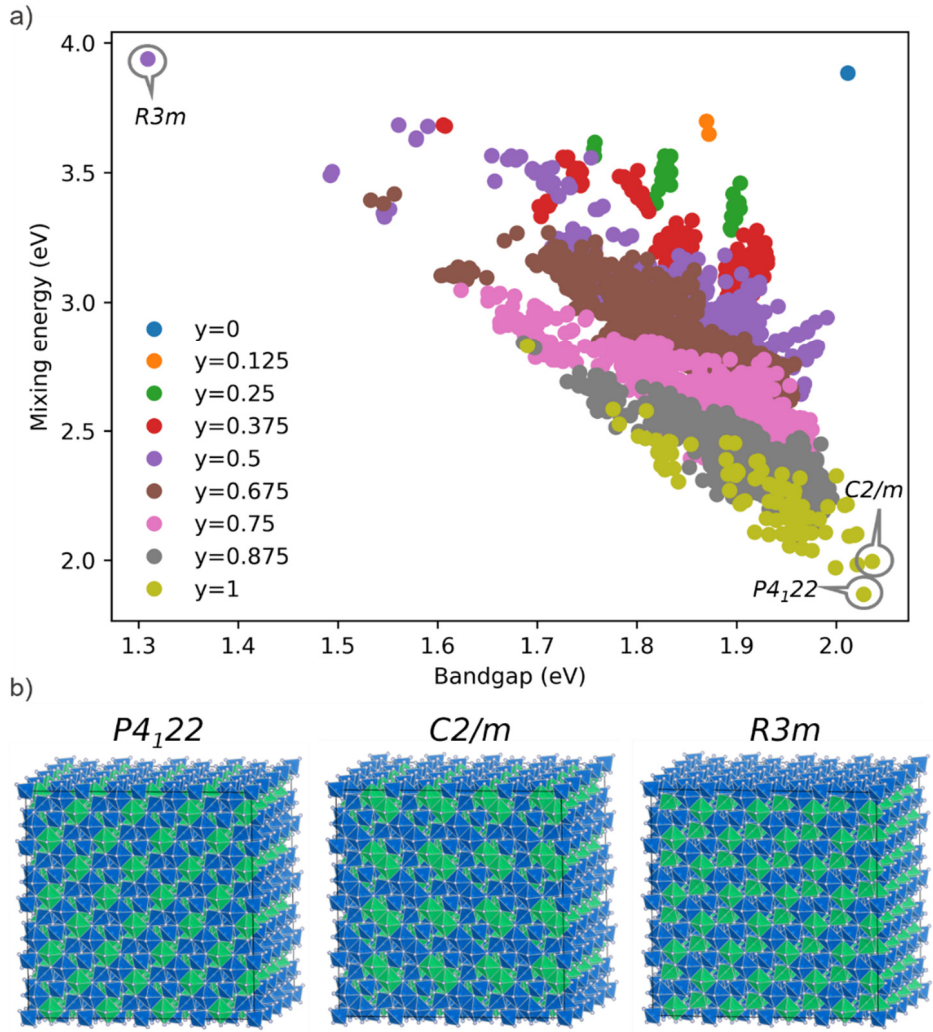

**Fig. 6** a) Distribution of predicted mixing energies and bandgaps for all the inequivalent Ge/Sn configurations in the GeSn$_2$N$_4$ cubic unit cell. Different colors represent different inversion degrees. b) Structures of extremal configurations: the P4$_1$22 configuration has the lowest energy, the C2/m configuration has the widest bandgap, and the R3m configuration has the narrowest bandgap but highest energy. Sn in blue, Ge in green.

A configurational map for the two studied properties is presented in **Fig. 6a**, giving us a broad perspective of the complete configurational space. This plot illustrates the mild inverse correlation between mixing energy and bandgap ($R^2$=0.405), which implies that there will be a thermodynamic preference for the widest bandgaps. The marked dependence of the mixing energy (but not of the bandgap) on the inversion degree is apparent from the plot. It also allows us to identify the extremal configurations, i.e., those with the highest and lowest bandgaps and mixing energies. Those configurations have been highlighted in the map and their structures, along with their space symmetry group, are also presented in **Fig. 6b**.

The most energetically favorable configuration, which also exhibits one of the widest bandgaps (2.03 eV) within the configurational space, belongs to the P4$_1$22 space group. This configuration is fully inverted, *i.e.,* it has tetrahedral sites fully occupied by Sn, whereas the octahedral sites contain both Sn and Ge atoms in an ordered pattern. This ordered configuration corresponds to the tetragonal structure of Zn$_2$TiO$_4$,[54] and is known to be the configurational groundstate for other inverse spinels.[55–58] It has two distinct octahedral sites, on which the two cations are ordered. Although the P4$_1$22 configuration has a very wide bandgap, the widest bandgap (2.04 eV) configuration is another fully inverted one with C2/m symmetry and a mixing energy not much higher than the P4$_1$22 configuration. Finally, the least stable configuration, with the highest mixing energy and the lowest bandgap (1.31 eV), is a structure with an inversion degree *y*=0.5 and R3m space group and has a quasi-2D structure with alternating layers of Sn and Ge cations.

The analysis above refers to the properties of individual configurations. However, given the small energy differences in the configurational space, we can expect that there will be a large degree of disorder in the GeSn$_2$N$_4$ structure, which we will discuss now based on the full configurational energy spectrum. For the determination of the equilibrium degree of inversion from first-principles calculations, it is common to adopt a simple 3-point model[30,59–61] based on the DFT energies of the primitive cell in its three possible degrees of inversion, *y* = 0, 0.5, and 1 (there is only one symmetrically different configuration of cations for each value of *y* in this cell). In this model the inversion energy, $E_{\text{inv}}(y)$, is a quadratic fit of the DFT energies, and the equilibrium degree of inversion at a given temperature *T* is given by the minimum of the inversion free energy:

$$F_{\text{inv}}(y, T) = E_{\text{inv}}(y) - TS_{\text{inv}}(y)$$

where

$$S_{\text{inv}}(y) = -k_{\text{B}}(y \ln y + (1-y) \ln(1-y) + y \ln\frac{y}{2} + (2-y) \ln(1-\frac{y}{2}))$$

is the "ideal" configurational entropy of inversion, assuming no energy differences between configurations with the same degree of inversion ($k_{\text{B}}$ is Boltzmann's constant). For our system, this model leads to the prediction of an almost fully inverted supercell at any temperature of interest, as shown in **Fig. 7**. The advantage of this model is that it does not require the evaluation of DFT energies in a large configurational space, but only three energies in a small supercell.

However, the underlying assumption that the energy is only a function of inversion degree is not correct for GeSn$_2$N$_4$, as has been demonstrated above. Having access to all the energies (and other properties) in the much larger configurational space for the cubic conventional cell allows us to perform more accurate configurational statistics. The equilibrium degree of inversion, for example, can be calculated as:

$$y_{\text{eq}}(T) = \sum_{m=1}^{4222} P_m y_m$$

where

$$P_m = \frac{1}{Z}\exp(-E_m/k_\text{B}T)$$

is the Boltzmann probability of configuration $m$, with energy $E_m$ as predicted from the ML model, and $Z$ is the partition function that guarantees that the sum of probabilities is one.[43,62] In contrast with the result from the 3-point model, the equilibrium degree of inversion calculated via configurational averaging departs significantly from 1. This means that the cation distribution in this system will depend on the thermal history of the sample. If the system is allowed to equilibrate at low temperatures (say if annealed very slowly), the inversion will be almost complete ($y \approx 1$). But if the synthesis procedure somehow freezes the high temperature disorder, say by rapid quenching after synthesis, an incomplete inversion might be achieved. Control of the cation distribution in this way is interesting and might have practical applications, because it provides a route to tune the bandgap, and perhaps other properties, of the system. In the low temperature limit, the bandgap equals the gap of the lowest-energy configuration, 2.03 eV, whereas in the high-temperature limit, with a random cation distribution corresponding to $y=2/3$ (*i.e.*, equal to the Sn/(Sn+Ge) ratio), the average bandgap reduces to 1.87 eV. These useful predictions require knowledge of energies and bandgaps in a large space of thousands of cation distribution configurations, so this type of analysis in other systems would be very computationally expensive if it was not accelerated by the ML techniques presented in this work.

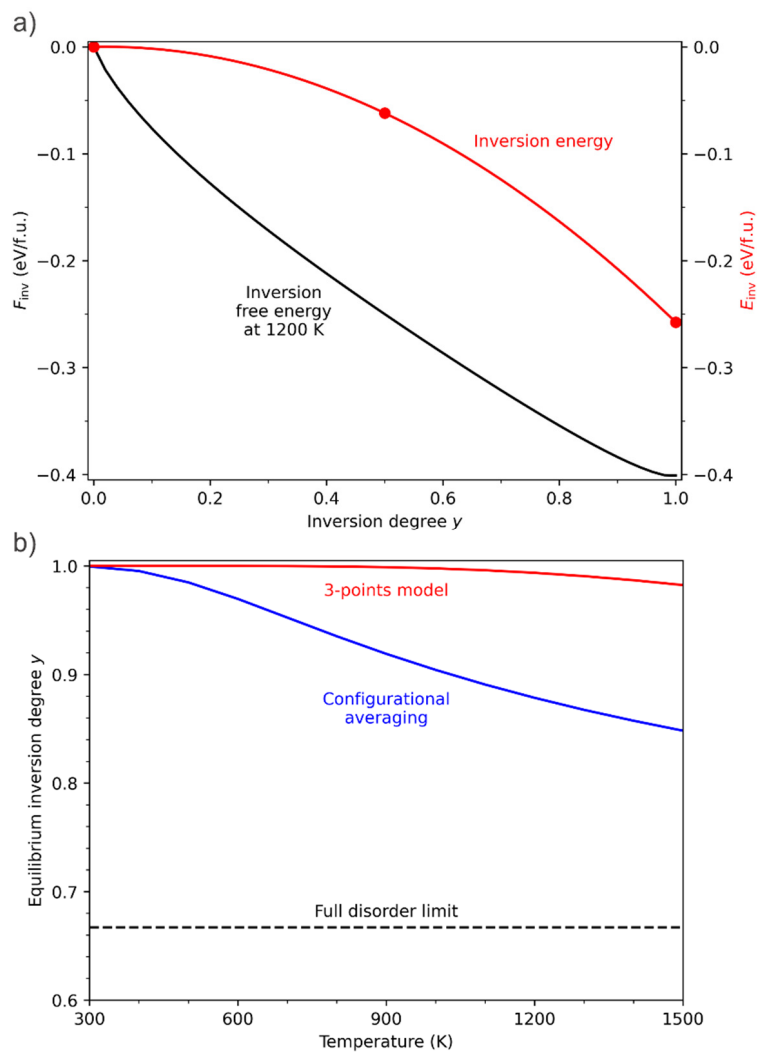

**Fig. 7** a) Inversion energy (in red) calculated with the 3-point model, and the corresponding free energy at 1200 K. b) Equilibrium degree of inversion predicted from averaging in the 4222-point configurational space (in blue), in comparison with the prediction from the 3-point model (in red) and the full-disorder limit (black dashed line).

## 4. CONCLUSIONS

We have presented an accurate and interpretable description of the whole configurational space of γ-phase $GeSn_2N_4$ nitrides, at a greatly reduced cost, through the combined use of DFT calculations and ML techniques. Our ML models, trained on DFT results from a small fraction (20% at most) of the structures within the space, exhibit excellent performance metrics, with mean errors in the range of few meV and almost perfect correlation between ML-predicted and DFT-calculated values.

Our results provide useful methodological information to perform this type of study in the future. We have compared the performance of different descriptors and models and found the optimal combinations for each task. In this case, a linear model based on cluster correlation functions, i.e., a cluster expansion, is shown to be the best model for the energies. For bandgap predictions in the configurational space, the non-linearity of the neural network (based on cluster correlation functions) has the best performance. Explainable ML highlights difference between energy and bandgap predictions, for the latter, the most relevant clusters are not necessarily the smallest and lowest-order ones, which means that cluster expansions of the bandgap, even non-linear ones, require a very large cluster basis. It is interesting, from the comparison with previous work, that the conclusions about the best descriptors and models are not universal. It is unclear so far how the optimal model depends on the nature of the solid solution, and this will be the subject of future research.

For the spinel nitrides considered here, we have seen that the configuration energies are mainly, though not completely, influenced by the inversion degree, and in second term by pair-type clusters. High inversion degrees, i.e., Sn occupation of tetrahedral positions, are clearly favored. However, we have demonstrated that the accurate calculation of the equilibrium inversion degree at a given temperature requires consideration of the energy differences at a given degree of inversion, thus illustrating the limitations of traditional equilibrium inversion models for spinels. Our combined DFT and ML model allows to predict that the bandgap of this solid solution can be potentially tuned via feasible modifications of the cation distribution in the system. The combination of methods that we have demonstrated and the insights that we have obtained should be applicable to many other alloy and solid solution systems.

## DATA AVAILABILITY

The code and data to reproduce the results in this paper are available at https://github.com/pablos-pv/GeSn2N4_ML

## DECLARATION OF COMPETING INTEREST

The authors declare that they have no known competing financial interests or personal relationships that could have appeared to influence the work reported in this paper.

## ACKNOWLEDGMENTS

This work was accomplished thanks to the mobility grants given by Universidad Politécnica de Madrid Programa Propio de I+D+I 2021 and ERASMUS+ European project. P.S-P. and P.P. acknowledge support from Ministerio de Ciencia e Innovación through the project BESTMAT-QC (PID2019-107137RB-C22). S.H. also acknowledges funding from the Agencia Estatal de Investigación and the Ministerio de Ciencia, Innovación y Universidades, of Spain (PID2019-110430G B-C22), and from the EU FEDER Framework 2014-2020 and Consejería de Conocimiento, Investigación y Universidad of the Andalusian Government (FEDER-UPO-1265695). The authors also gratefully acknowledge the UK Materials and Molecular Modelling Hub, which is partially funded by EPSRC (EP/T022213/1), for providing computing resources on the Young supercomputer.